\documentclass[rms3]{article} 
\usepackage{epsf} 
\usepackage{graphics}
\usepackage{manuscript}
\usepackage[round]{tellus}  
\begin{document} 

  \ijc{Benestad}
    {Trend in number of tropical cyclones}
    {2005}
    {arXiv}
    {1}
    {14}

\bibliographystyle{tellus}

\title{\sc  An explanation for the lack of trend in the hurricane frequency}

\author{Rasmus E. Benestad\\
        Norwegian Meteorological Institute, PO Box 43, 0313, Oslo, Norway\\
        rasmus.benestad@met.no}

\date{Submitted: \today}

\summary{
The proposition that the tropical cyclogenesis increases with the size of the warm pool, the area enclosed by the 26\degC\ SST isotherm, is tested by comparing the seasonal variation of the warm pool area with the seasonality of the number of tropical cyclones. A non-linear relationship of high statistical significance is found between the area and the number of cyclones, which may explain both why there is no linear trend in the number of cyclones over time and the recent upturn in the number of Atlantic hurricanes. 
}

\clearpage

\section{Introduction}

After active hurricane seasons in the Atlantic/Carribean basins in 2004 and 2005, the awareness about tropical cyclones grew in the media. Furthermore, the hurricanes sweeping across the Carribean affected US oil production, having ring effects world wide. Thus, indirect effects from the hurricane activities could be felt as far afield as Norway. One frequent question was: is the hurricane activity affected by the global warming?  
High numbers of Tropical cyclones (TCs) in the Atlantic and Caribbean during 2004 (15 named TCs) and 2005 (27 named TCs, three of which were category-5, four landfalls caused extensive damages, and Wilma reached the highest intensity on record in the Atlantic basin\footnote{http://en.wikipedia.org/wiki/Hurricane\_Wilma}) have spurred further speculation of whether TCs can be affected by a global warming \citep{Trenberth05,Scharroo2006,Sun2006,Michaels2005,Knutson2005,Pearce2005b,Smith2005}. There have been reports of increased activity (e.g. number of TCs) in the period 1995--2000 compared to 1971--1994 due to increases in the North Atlantic sea surface temperature (SST) and decrease in wind shear \citep{Goldenberg2001}. Goldenberg {\it et al.} posed the question whether the increased activity was due to long-term global warming or a result of natural variability, but concluded that the latter was the most likely explanation. They argued that the recent high north Atlantic SST and reduced vertical shear will persist for some years to come, and predicted that the high activity is likely for subsequent 10--40 years. Three more recent studies have suggested that the potential destructive energy in TCs has increased since 1970 \citep{Emanuel05} and that the number of intense cyclones has risen over the time period \citep{Hoyos2006,Webster05}. However, the question whether there is a systematic relationship between the area of high SST and number of TCs is still debated. Furthermore, a systematic change in the total number of TC has not yet been established and a lack of trend may seem contrary to expectations given a general warming trend. 

Details of physical conditions and processes associated with TC formation and their intensity are poorly known. One pressing question is whether  TCs will become more frequent in the future or more intense under a global warming as some model results indicate an increase in intensity and near-storm precipitation rates with $CO_2$-induced warming \citep{Knutson2004}.
Global climate models (GCMs) provide an important tool for making future climate scenarios, but these do not yet have a sufficient spatial resolution or a representation of the physical processes within the storm to give accurate results \citep{Henderson-Sellers98,Jung2006b}. It is nevertheless well-known that the formation of tropical cyclones requires SST greater than 26--27$^\circ$C, whereas strong vertical wind shear in the troposphere provides unfavourable TC conditions \citep{Goldenberg2001,Pearce2005,Henderson-Sellers98}. 
A remote sensing study of hurricane Katarina-2005 suggests that the rapid increase in intensification of the hurricane was more likely due to oceanic dynamic topography, representing the upper ocean heat content, rather than SST \citep{Scharroo05,Scharroo2006}. A stirring of the upper ocean layers by the high winds associated with TCs bring subsurface water up to the surface, and high SSTs is not sufficient as a deep thermocline is required to maintain high SSTs during the mixing or Ekman pumping caused by hurricanes. However, high SSTs are often associated with a thick upper layer of warm water (a deep thermocline).

\cite{Henderson-Sellers98} argue that it is a widespread misconception that the tropical cyclogenesis increases with the area enclosed by the 26\degC\ SST isotherm and base their statement on an application of a thermodynamic technique to climate change scenarios. However, the thermodynamic technique cited by \cite{Henderson-Sellers98} is tailored for the intensity of TCs rather than their frequency. The statement about the relationship between the warm area and cyclogenesis is re-examined in the present analysis through a different data analysis approach, as \cite{Henderson-Sellers98} do not provide convincing evidence for why the cyclogenseis should not be sensitive to warm pool area.
Here only the frequency of tropical cyclones is examined, but the number of cyclones does not necessarily give an adequate indication of the severity of the tropical cyclone activity, as aspects such as trends in individual cyclone life times, intensities \citep{Sun2006}, and spatial size have not been included in this simple analysis.

\section{Methods \& Data}

\subsection{methods} 

The objective of this analysis was to test the hypothesis of a systematic relationship between the number of TCs and the area of the region where SST is greater than 26.5$^\circ$C (here represented by the symbol $A$).
If it can be assumed that there is no systematic change in the atmospheric conditions, that the TCs are independent of each other, and TC-formation can be represented by a stochastic process, then the probability of observing a TC can be expected to be proportional to $A$:

\begin{equation}
Pr(\mbox{TC} | A) \propto A,
\end{equation} 

\noindent and the number of TCs, $N$, is proportional to the area. Since the TCs may disturb their own environment (winds, and SST), they are strictly not independent. It is possible that they are clustered in time as a result of weak interactions and non-linear behaviour. For instance, the convection associated with TCs may act to maintain low vertical wind shear by equalising upper and lower level horizontal momentum, but TCs also remove heat from the ocean surface through their action of vertical redistribution of heat. One may nevertheless expect an approximate number of TCs to be proportional to $A$ if the probability of an event is low and few TCs coincide in time and space. 

\subsection{Data} 

Here the data on Atlantic/Caribbean (henceforth referred to as ``Atlantic'') TCs (1851--2004) was taken from the National Hurricane Center \footnote{http://www.aoml.noaa.gov/hrd/hurdat/hurdatTAB.txt}, but the TC data for the northwest Pacific (1950--2003) and Indian Ocean (1971--2002) were taken from US Navy best-tracks\footnote{http://metoc.npmoc.navy.mil/jtwc/best\_tracks/}.
The SST was the NOAA extended reconstruction from NOAA CDC\footnote{http://lwf.ncdc.noaa.gov/oa/climate/research/sst/sst.html}. The GCM data was produced by HadCM3 \citep{HadCM3}: a 20th Century run for the past and the IPCC SRES A1b emission scenario for the future\footnote{https://esg.llnl.gov:8443/index.jsp}.

\section{Results}

\begin{figure}[t]
\centerline{
\resizebox{8.0cm}{8.0cm}{
\includegraphics*{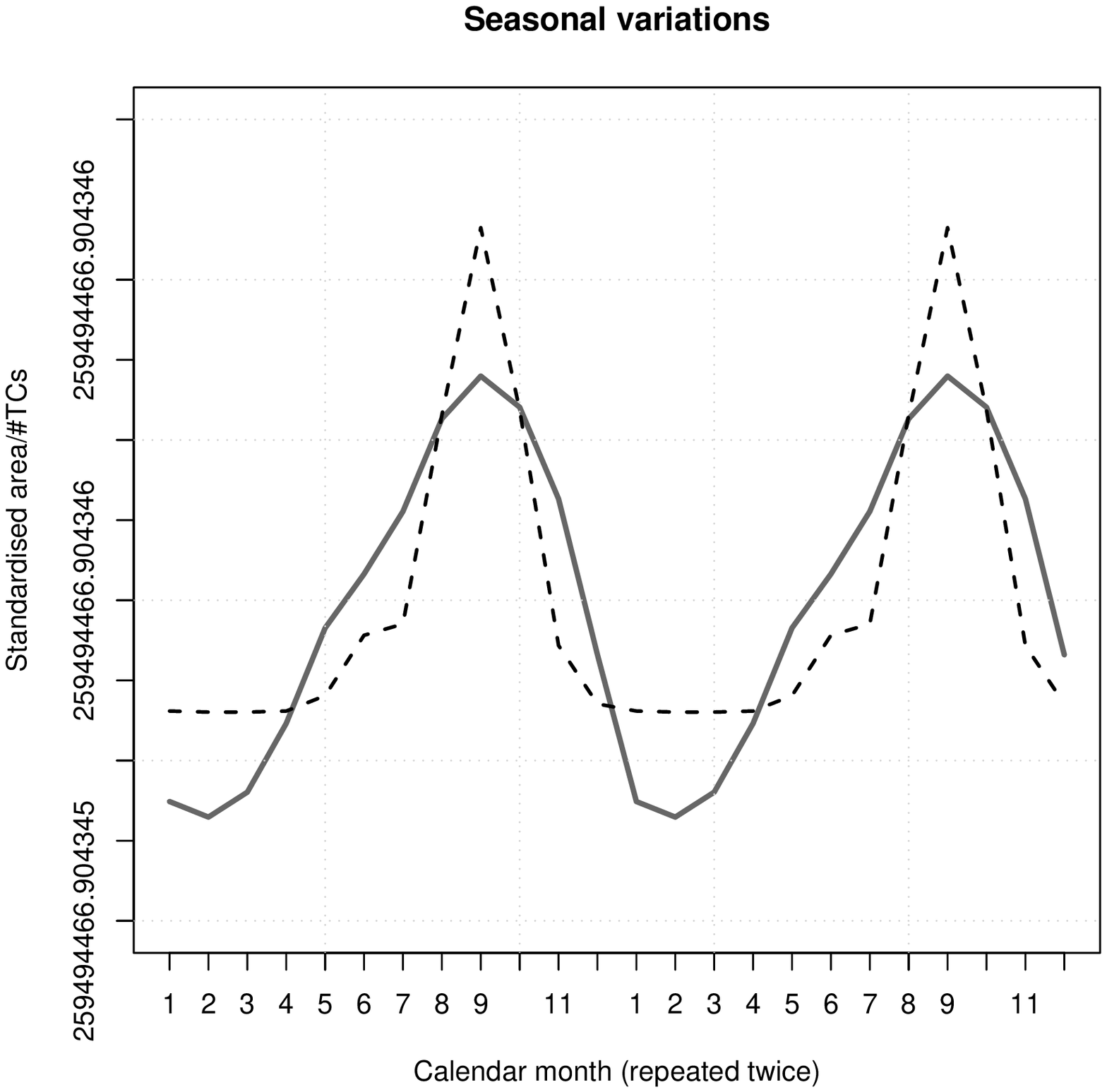}}
\resizebox{8.0cm}{8.0cm}{
\includegraphics*{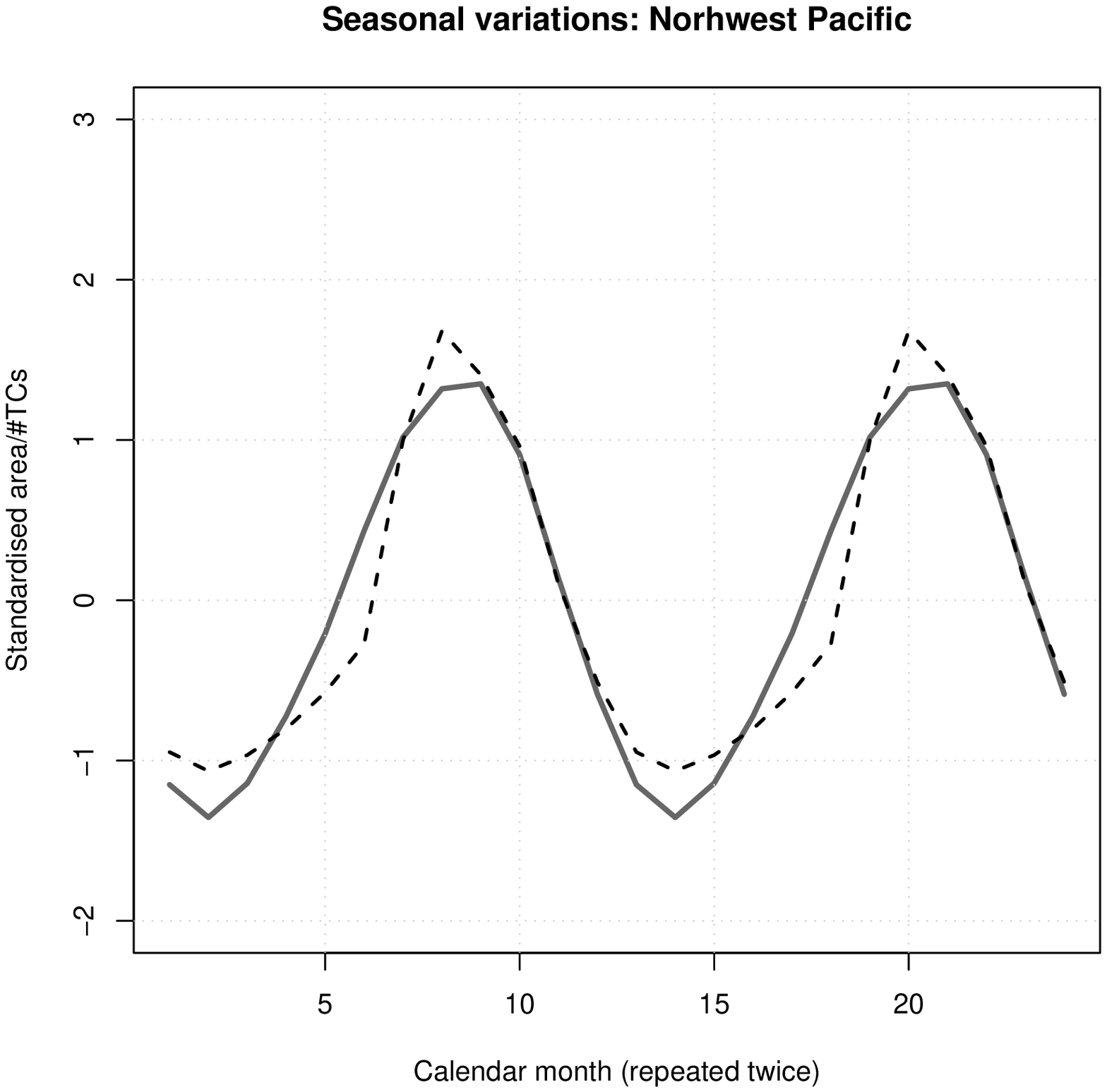}}
}
\centerline{
\resizebox{8.0cm}{8.0cm}{
\includegraphics*{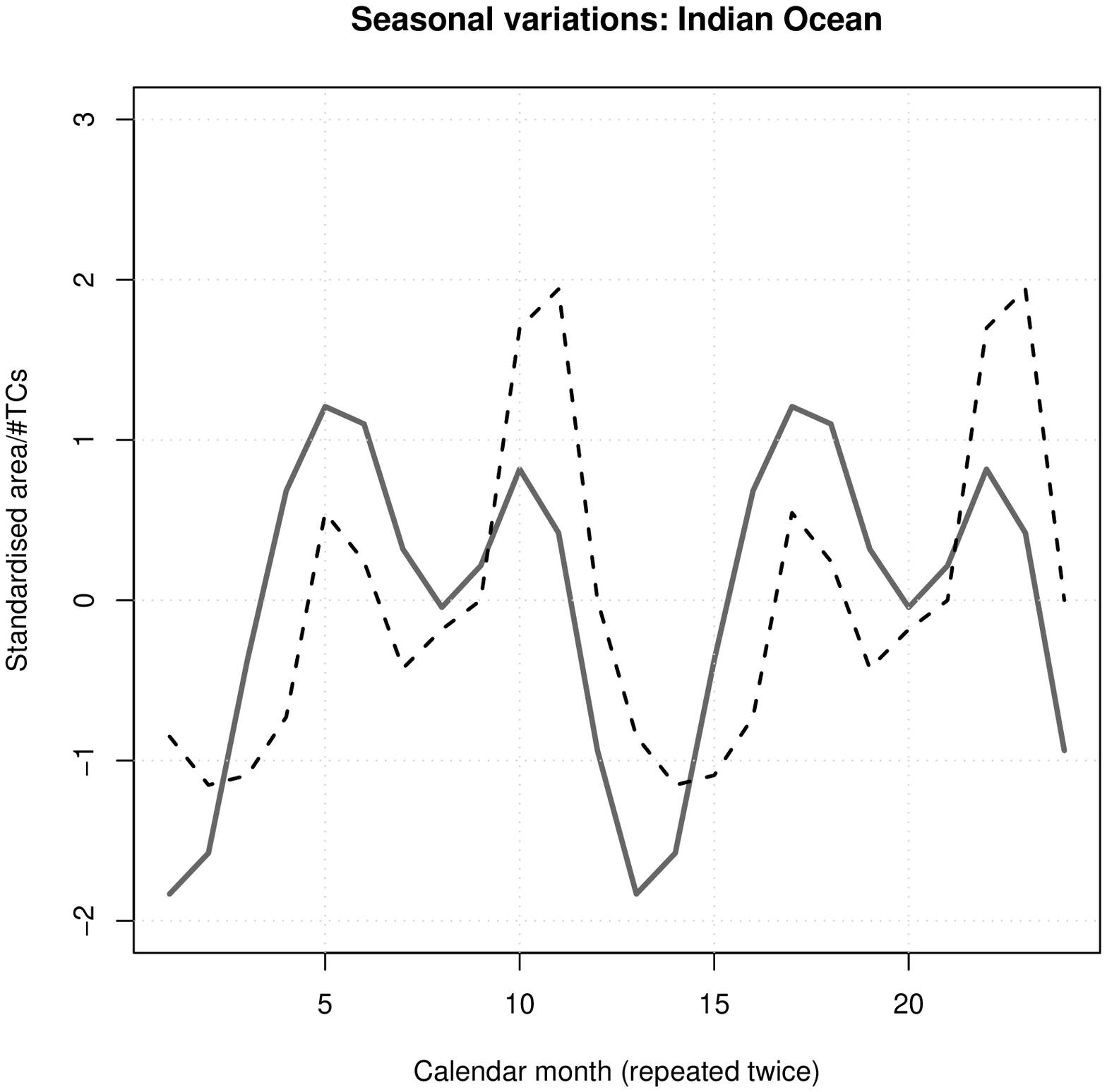}}
}
\caption{\small
The annual variation in surface area of SST $>$ 26.5$^\circ$C\ ($A$; grey) and the number of TCs ($N$; dashed) for the (a) Atlantic/Caribbean basin, (b) the northwest Pacific, and (c) the Indian Ocean. All the curves have been standardised.
}
\label{fig:1}
\end{figure}

The annual cycle in the TCs, $N$, are shown in Figure \ref{fig:1} as black dashed line whereas the grey curves represent the annual variation in $A$. 
A comparison between the annual cycle of the area of SST exceeding a critical temperature (here taken as 26.5\degC) in the North Atlantic (henceforth referred to as 'the warm area') and the seasonal variation in the Atlantic TC number both show a clear 12-month annual cycle (Figure \ref{fig:1}a). 
Subscripts, e.g. $N_{Atl}$, $N_{Pac}$, and $N_{Ind}$, are henceforth used to indicate the region represented by the data/analysis whereas symbols with no subscripts are used for more general discussion.

An interesting observation is that there is not a one-to-one ratio between $N$ and $A$. For the Atlantic region (here $A_{Atl}$ was estimated over the region -80--10$^\circ$E / 0--40$^\circ$N), there is a disproportionally high number for the month with greatest area. Thus, the hypothesis that the number of TCs is proportional to the warm area and the number of TCs therefore appears to be inconsistent with these results. For the northwest Pacific (here 100$^\circ$E -- 150$^\circ$W / 0--30$^\circ$N), on the other hand, the annual cycle of $N_{Pac}$ and $A_{Pac}$ exhibits more of a linear relationship, however, the peak in TC number is still narrower than that of the warm area. Over the Indian Ocean (here 40--120$^\circ$E / 0--30$^\circ$N), the seasonal cycle is characterised as a double peak in both $A_{Ind}$ and $N_{Ind}$, indicative of a real relationship between the two. However, the second peak in $N_{Ind}$ is more pronounced than the first peak whereas for $A_{Ind}$ the first peak is more prominent. 

The relationship between the warm surface area and the number of cyclones can be explored further through more sophisticated statistical analysis. In order to reduce the influence of other factors affecting the signal-to-noise ratio, the mean seasonal cycle, rather than the individual months, was used for developing a statistical model for the relationship between number of TCs and the warm area. If the effect by other influences (e.g. the noise) follows a Gaussian distribution it will tend to cancel when taking the average over a long interval, {\em if} these were unrelated to the seasonal cycle (such as El Ni\~{n}o conditions) or the SST itself. Furthermore, the low number of TC-events for each month or each season, which in reality reflects a low probability $Pr(\mbox{TC} | A)$, hampers the attribution analysis. An average over longer interval improves the statistical power, but the question concerning whether the calibration is biased by other factors also exhibiting an annual cycle but not related to SST also has to be addressed.

%

\begin{figure}[t]
\centerline{
\resizebox{12.0cm}{12.0cm}{
\includegraphics*{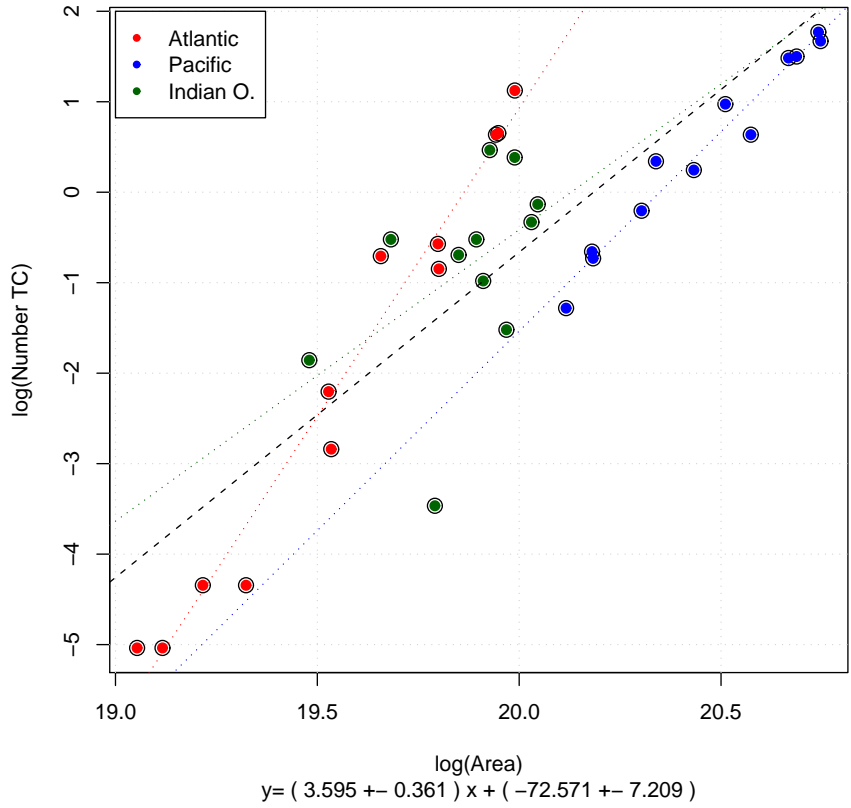}}
}
\caption{\small
Scatter-plot of seasonal values x=log(A) and y= log(N). Here $A$ is in units of km$^2$ and $N$ in number/month. 
}
\label{fig:2}
\end{figure}

The logarithm of the seasonal variation in warm Atlantic surface area ($x=\log(A)$) is compared with the logarithm of the seasonal cycle in monthly mean number of Atlantic TCs, $N_{Atl}$ ($y=\log(N)$), and the relationship between $x$ and $y$ has a predominately linear character (Figure \ref{fig:2}; red symbols) with a linear slope suggesting $N_{Atl} \propto A_{Atl}^{7.0 \pm 1.3}$ ($\pm$ 2 standard deviations). The adjusted R-squared statistic for this relationship was 0.9382, with an F-statistic of 107.3 on 1 dependent variable and 6 degrees of freedom (DF; here only the TC season is included) and a p-value of $5 \times 10^{-05}$. In other words, the linear relationship between $x$ and $y$ is highly significant. The linear relationship can be compared with those of the two other ocean basins (blue and green symbols in Figure \ref{fig:2}), and the data representing the northwest Pacific indicates similar linear relationship between the $x$ and $y$, but the slope for the northwest Pacific is weaker: $N_{Pac} \propto A_{Pac}^{4.4 \pm 0.5}$ (Adjusted R-squared= 0.9584, F-statistic= 254.6 on 1 and 10 DF, \& p-value= $2 \times 10^{-08}$). The relationship over the Indian Ocean is weak despite the annual double peak, and $N_{Ind} \propto A_{Ind}^{3.2 \pm 3.9}$ (Adjusted R-squared= 0.145, F-statistic= 2.696 on 1 and 9 DF, and  p-value= 0.135). Both the Indian Ocean and northwest Pacific records are short compared to the Atlantic record, and the annual cycles are hence more strongly affected by noise. Although the relationship between $x$ and $y$ for the Indian Ocean was characterised by large scatter, their relationship fit well in the general linear relationship for all ocean basins suggesting $N \propto A^{3.6 \pm 0.7}$ (black dashed line in Figure \ref{fig:2};  Adjusted R-squared= 0.9662, F-statistic= 315 on 1 and 10 DF, and  p-value= $7 \times 10^{-09}$). One reason for the weak relationship between $x$ and $y$ over the Indian Ocean, and the different magnitudes of the annual double peak, may be that effects of atmospheric conditions on TCs are more pronounced or that the shape of region matters for TC development.

The Atlantic TC data after 1944 is thought to have higher quality than in the earlier observations \citep{Goldenberg2001}, and a linear relationship was derived between the seasonal cycle taken over the interval 1944--2004. Regression results for $\hat{y}_m = \alpha x_m + \beta$, where $m$ represents the different months in the TC season, yielded the empirical expression for the mean number of Atlantic TCs in terms of the Atlantic warm region area: 

\begin{equation}
N_{Atl} \propto A_{Atl}^{6.64 \pm  1.10}.
\label{eq:atl_1944-2004}
\end{equation}

\noindent The linear least-squares regression analysis suggested that the p-value for this relationship is of the order $10^{-5}$, $R^2$=0.96, and F-statistic of 148 on 6 degrees of freedom. 

So far, the possibility that other factors important for TCs also exhibiting an annual cycle has not been ruled out. One way to isolate and assess the importance of warm sea area with respect to $N$ is then to apply the statistical relations derived above to predict year-to-year variations in the seasonal mean number of TCs over an interval representing independent data, and subsequently evaluate against the observations. This approach is similar to one used by \cite{Michaels2005} to assess the association between SST and the total number of annual hurricanes, however, here the SSTs (in their analysis averaged over 10\degN--25\degN\ and 15\degW--80\degW) were substituted with the predicted values using expression \ref{eq:atl_1944-2004} (note, the conclusions drawn here contrast with those made by Michaels {\em et al.}).
When this model was applied to mean $A_{Atl}$ corresponding to the hurricane season (June--November) of each year over the 1944--2004 interval, a correlation of 0.43 (p-value=0.0005, assuming independent and identically distributed data) was achieved. For the older data (representing 1851--1943) with presumed lower quality, a weaker correlation (0.27) was found, but it was still statistically significant at the 1\% level (p-value=0.009). In other words, the empirical expression captures some of the year-to-year variations in the TC numbers over the independent evaluation period. On the other hand, a similar correlation analysis for the Northwest Pacific over the independent years 1950--1987 yielded a low correlation (r=0.14) with a p-value of 0.4137, and a negative correlation (r=-0.22) for the Indian Ocean over the interval 1971--1992 (p-value = 0.3207).

\section{Discussion}

It cannot be ruled out here that the correspondence between $A$ and $N$ seen in Figures \ref{fig:1} and \ref{fig:2} may be due to other factors than SST, as atmospheric conditions important for TC generation may also be affected by the annual cycle in a similar was as the SST. However, the independent correlation analysis carried out for the seasonal totals above for the Atlantic/Carrebian basin suggest statistically significant results, and both $N_{Ind}$ and $A_{Ind}$ exhibit double peaks despite negative correlation for the seasonal totals. Further work is required to discriminate the role of atmospheric processes and establish whether atmospheric features important for TCs over the Indian Ocean follow similar bimodality.  
The statistical models trained on seasonally varying values did not yield skillful predictions for year-to-year variations in $N$ for all ocean basins. The reason for the negative correlation between predicted and observed year-to-year variations in $N_{Ind}$ may be associated with the different magnitudes in the double-peak structure, the weak statistical relation, and the large scatter in seen in Figure \ref{fig:2}. The low year-to-year correlation and low statistical significance for the Indian Ocean and the northwest Pacific may also be a result of lower data quality in these basins, shorter series, or due to stronger influence from other factors such as atmospheric conditions not directly related to the warm pool area. 

There are further limitations to the data on which this analysis rests, as the TC series should not be considered homogeneous, since the ability to detect TCs in the open Atlantic has increased substantially over time as aircraft reconnaisance and (in the 1970s) satellite monitoring have become available.  These improvements in detection tools have led to enhanced probability of detection of weak and remote TCs over time, although estimated maximum potential intensities of tropical cyclones appear to show some agreement with the observations \citep{Henderson-Sellers98}.
Using the seasonal variations in $A$ and $N$ defined over the 1944--2004 interval to some extent alliviates problems associated with inhomogeneities in the TC record. The fact that the correlation analysis between predictions based on equation \ref{eq:atl_1944-2004} and actual observations yielded results significant at the 1\% level for {\em independent} (older) data, provides strong support for the statistical model established here for the Atlantic/Carribean basin.   

The warm area cannot account for all the variability and other factors, such as atmospheric conditions, also affect the number of TCs. An intriging question is whether the annual variation of such factors are independent or affected by the warm area. 
Increases in the convective available potential energy (CAPE) are associated with increased near-surface temperature \citep{Gettleman2002}, suggesting that increased warm area may enhance convection over a greater region and hence cause a more widespread vertical equalisation of horizontal momentum, and thus act to reduce the vertical shear. 
Thus, the role TCs play in the vertical redistribution of momentum and hence their effect on the ambient atmosphere, may enhance the conditions of TC formation and growth. 
It is therefore plausible that the TCs are organised in time clusters, where the presence of one TC create conditions that may favour the genesis of other subsequent TCs, given sufficiently large area over which they can form. It is also plausible that a vertical mixing of heat and moisture may, on the other hand, act to inhibit further TCs.
Another speculation is whether the time clustering of TCs may be associated with a modulation of TC occurrence by the Madden-Julian Oscillation (MJO), or conversely that the MJO is affected by the TCs.  

Present analysis may point to anexplanation for why linear trends in the TC frequency has not been detected in the past. We can also carry the analysis further if a number of assumptions are made.

\subsection{A conditional scenario for the future, assuming warm area importance}

\begin{figure}[t]
\centerline{
\resizebox{8.0cm}{8.0cm}{
\includegraphics*{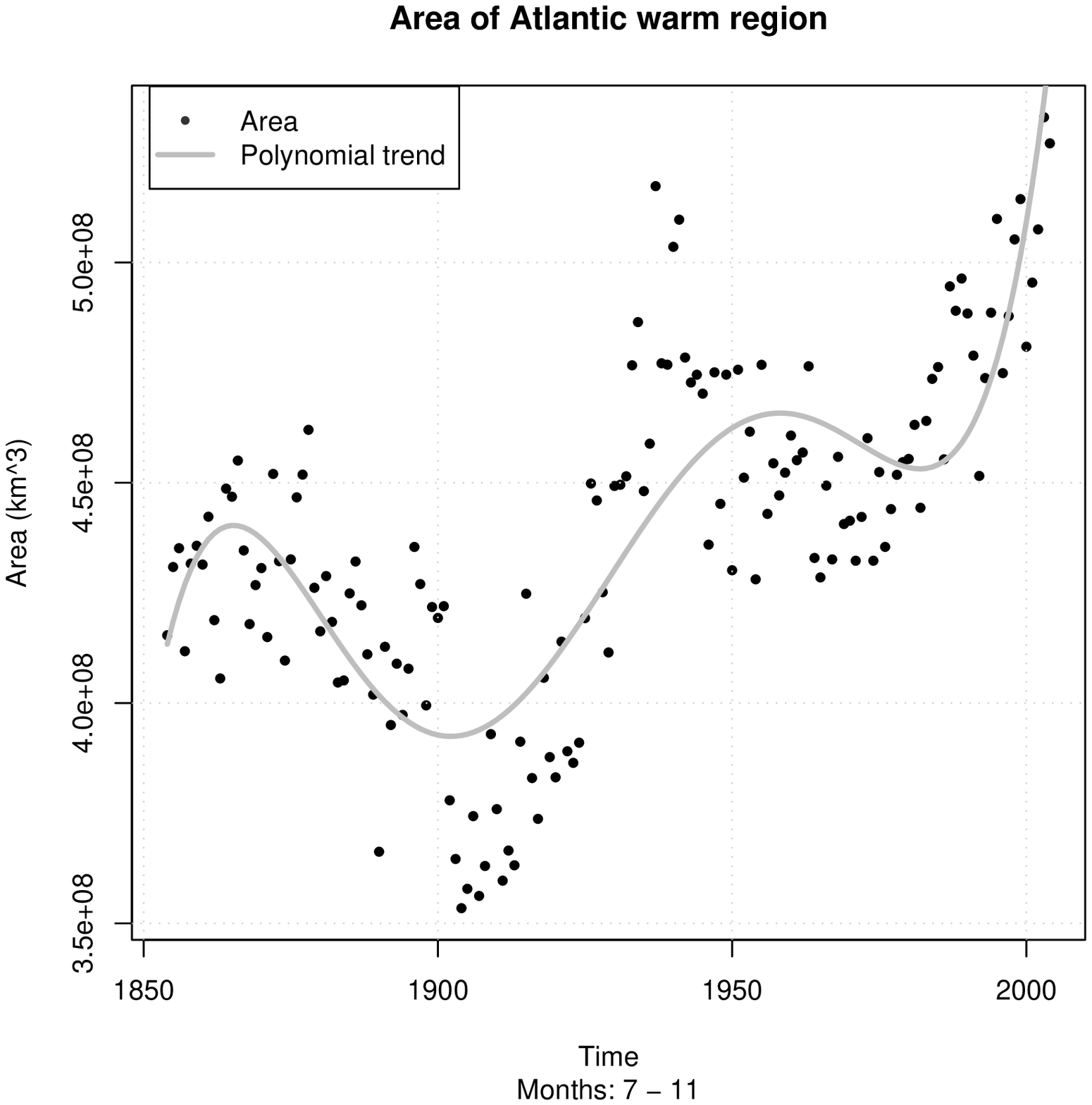}}
\resizebox{8.0cm}{8.0cm}{
\includegraphics*{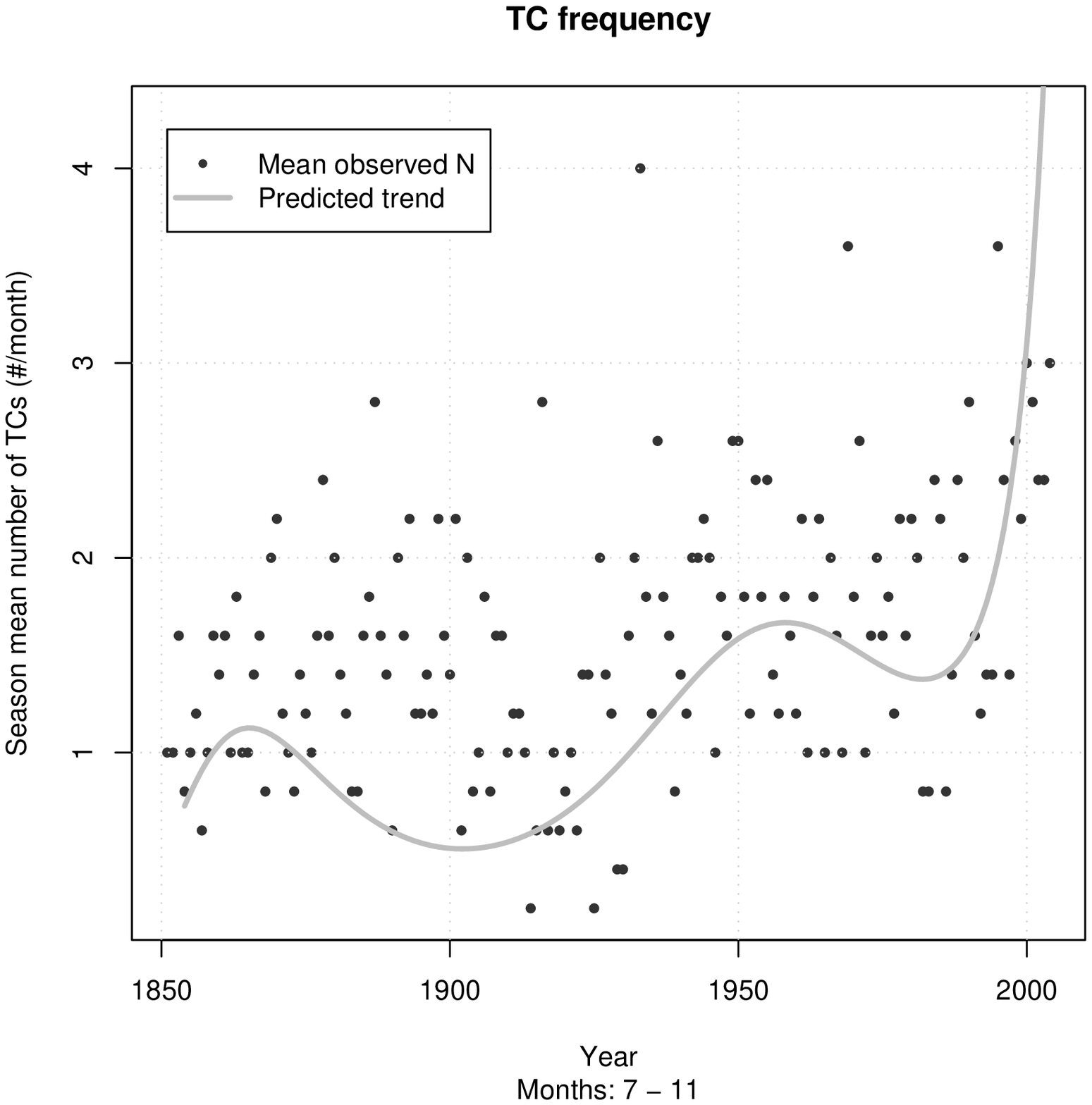}}
}
\caption{\small
(a) The area of the Atlantic warm region $A$ and (b) a reconstruction of the trend in $N$ for historic Atlantic TCs based on trend in $N_{Atl} \propto A_{Atl}^{6.64}$. Here a polynomial trend model was used because of the non-linear relationship between $A_{Atl}$ and $N_{Atl}$. Here $A$ is in units of km$^2$ and $N$ in number/month. 
}
\label{fig:3}
\end{figure}

If it can be assumed that the relationships between $A_{Atl}$ and $N_{Atl}$ in Figures  \ref{fig:1} and \ref{fig:2} and equation \ref{eq:atl_1944-2004} are not significantly affected by other factors \citep{Hoyos2006} also responding to the annually varying solar insolation, and these other factors being equal, then it is possible to infer trends for the future given changes in the warm area. Since it is unlikely that only the warm area is changing, the following scenario can only be regarded as a loose speculation. However, this exercise may nevertheless be useful for illustrating how dramatic the frequency of TCs may change given a non-linear dependency between TCs and the ambient climate. It is most likely wrong to assume a linear trend for the number of TCs, since we know there is a transition between TC occurrence at the critical temperature of 26.5\degC. Analogously, there is no {\em a priory} reason to assume that the evolution of local or regional temperatures to be linear \citep{Benestad02d}. Moreover, a non-linear relationship may account for a lack of trend in the historical record \citep{Goldenberg2001,Chan04} as the weak SST trend so far (Figure \ref{fig:3}a) has not yet lead to a large response in the frequency (Figure \ref{fig:3}b). High values in $A_{Atl}$ during 1930--1950, probably a consequence of natural variations, can explain the active hurricane season during those decades.
The recent upturn in the TC number, however, may be the beginning of more frequent TCs as the tropical temperatures slowly rise, however, on shorter time scales natural variations are also expected to modulate $N$. This upturn is also predicted when equation \ref{eq:atl_1944-2004} is applied to each individual TC season (not shown) as well as the trend analysis shown in Figure  \ref{fig:3}. The data therefore suggests that the linear log-log relationship holds for the range of values observed in the historical data.

\begin{figure}[t]
\centerline{
\resizebox{8.0cm}{8.0cm}{
\includegraphics*{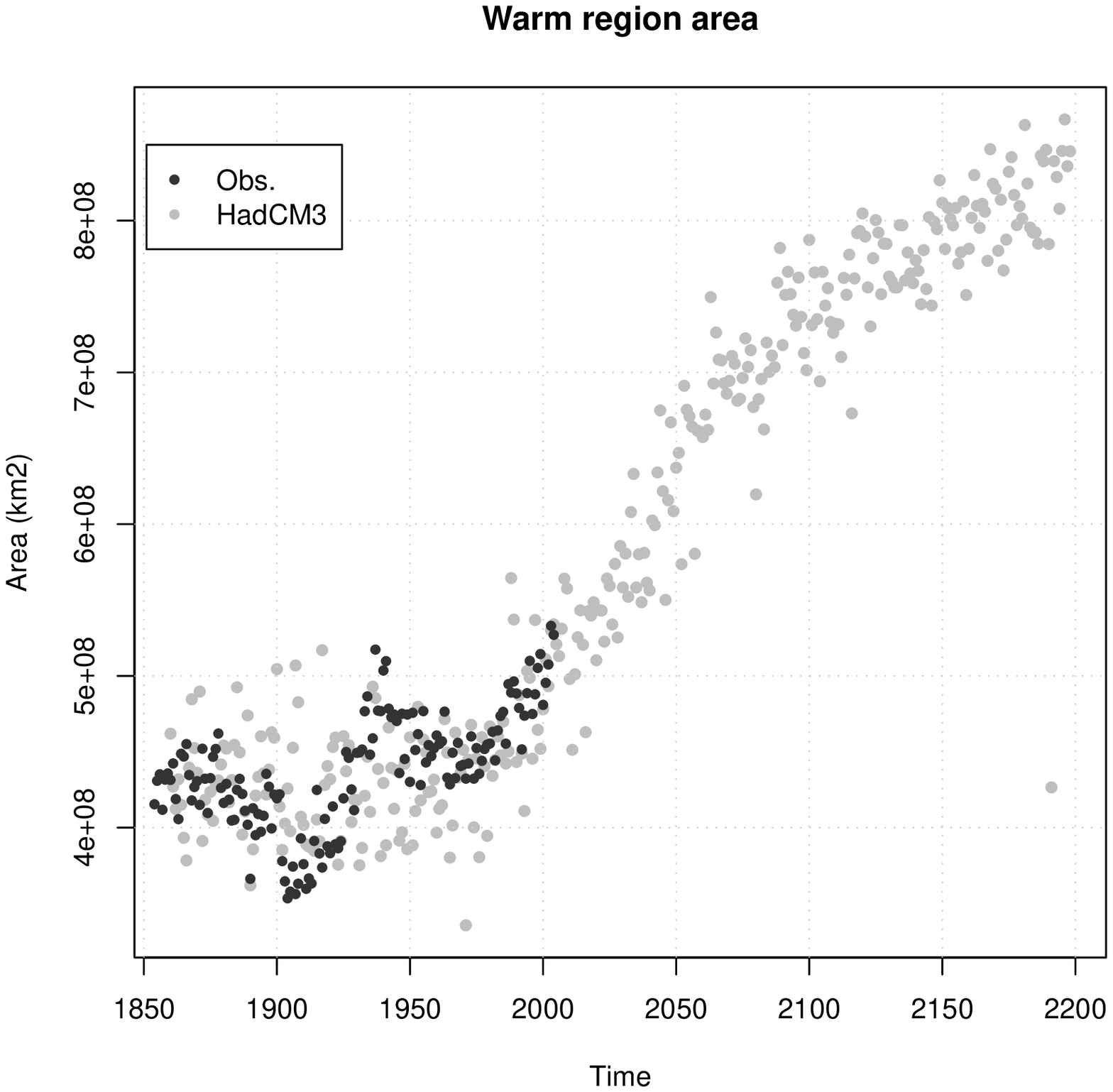}}
\resizebox{8.0cm}{8.0cm}{
\includegraphics*{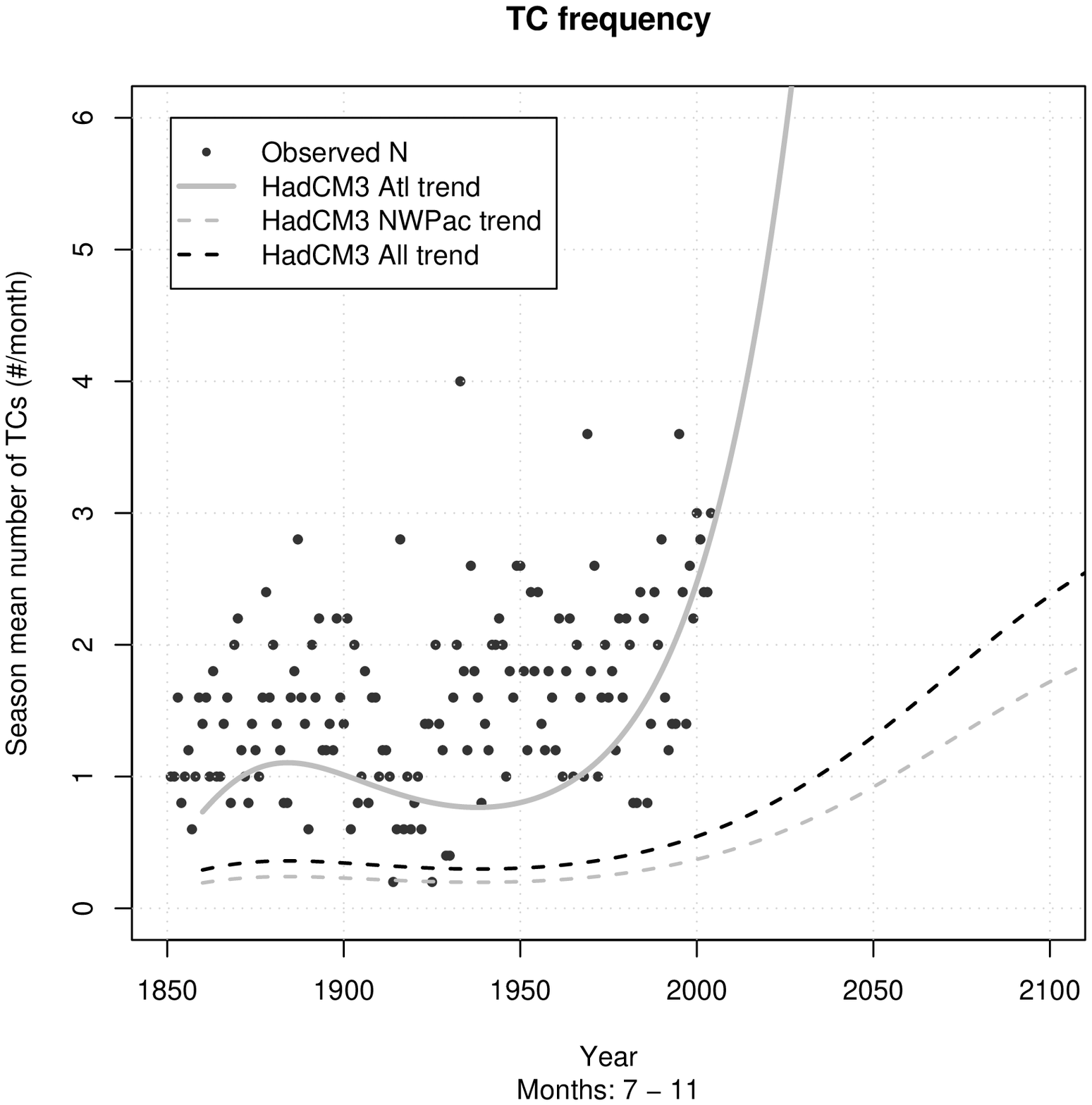}}
}
\caption{\small
A projection of the trend in Atlantic (a) $A_{Atl}$ and (b) $N_{Atl}$ based on HadCM3 results and $N_{Atl} \propto A_{Atl}^{6.64}$. The warm area in the GCM has been adjusted so that simulations of past area is at the same level as the observations as well as having similar standard deviations. Here $A$ is in units of km$^2$ and $N$ in number/month. 
}
\label{fig:4}
\end{figure}

A non-linear relationship between $N$ and $A$ is also concerning in the context of a global warming, as the implication of larger warm area in the future (Figure \ref{fig:4}a) would be greater numbers of TCs (Figure \ref{fig:4}b; red curve). However, it is not certain that the established relationships will hold for values beyond the range from which they were derived, as it is plausible effects of limiting factors may kick in for larger values of $A$. The weaker slope of the northwest Pacific in Figure \ref{fig:2}, may suggest a lower sensitivity to $A$ of greater values. If the slope $\alpha$ diminished with $x$, then a future change in $A$ is less dramatic than if $\alpha$ is constant. It is important to stress that the projection only should be regarded as an estimate of the upper limit on the TC-frequency in a situation where great uncertainties are associated with the relationship between TCs and the ambient environment \citep{Scharroo2006,Sun2006,Michaels2005,Knutson2005,Pearce2005b,Smith2005}. Hence, Figure  \ref{fig:4}b could be regarded as a 'worst case' scenario based on a combination of the analysis of empirical data and a global climate model projection of the warm sea area. 
Furthermore, $A$ is limited to the size of the ocean basin and has an upper limit. 

The trend analysis was repeated using the relationship established for the Pacific, $N \propto A^{4.4}$, (blue line) as well as for all basins, $N \propto A^{3.6 \pm 0.7}$, (black dashed) to explore the sensitivity of the results to the power of $A$. Since $A_{Atl}$ is small compared to that of the Pacific, the predicted $N$ using the Pacific relation is unrealistically low for the past. Both the expressions $N \propto A^{4.4}$ and $N \propto A^{3.6 \pm 0.7}$ suggest more moderate increases as a result of a future warming, however, since both underestimate the present-day number of TCs, it is questionable whether they are representative for the Atlantic. 
The major weakness of the projection is that it uses statistical relations developed for the present-day climate to extrapolate TC occurrence for future climate changes. 
\cite{Henderson-Sellers98} argued that elementary applications of empirical relationships from current climate to a future climate are fraught with danger and offer little useful insight.
It is likely that a greenhouse warming at the surface is expected to be accompanied by enhanced warming of the upper Tropical troposphere \citep[p. 544, Fig. 9.8.a]{IPCC2001c}.  Enhanced warming aloft may have an important effect on tropical cyclone intensification which is included in  potential intensity theories and dynamical modeling studies using hurricane models, but is an effect that a statistical analysis of seasonal variations is unlikely to account for \citep{Henderson-Sellers98}.
It should be stressed that the future scenario presented here can only be regarded as an estimate of the upper limit of number of TCs associated with increased warm pool size. This analysis also highlights the need for further work required to establish whether other factors have 'contaminated' the seasonal correlation between the warm area and TC frequency and may play a role for the future trends. 

The fact that the slopes in Figure \ref{fig:2} are greater than unity may suggest that favourable atmospheric conditions may be more frequent with greater warm area.
Yet, another question is whether processes that may trigger TC genesis are unaffected by the warm area and hence serve as a limiting factor for the number of TCs. Another posibility is that the probability is not constant above the critical temperature, but increases with SST. 
On the other hand, it may be argued that some global modeling studies indicate that the region of strong tropical convection, and of tropical storm formation, does not simply expand with some given isotherm (e.g., 26\degC) but rather the "critical temperature" increases to a higher value in the uniformly warmer climate (e.g., $\sim$29\degC\ for a 3\degC\ increase in the global mean surface temperature). \cite{Henderson-Sellers98} argue that the region of cyclogenesis will not expand with the 26\degC\ isotherm.

In this analysis, the distinction between the absolute value of SSTs and SST gradients has been neglected. If the SST is elevated uniformly everywhere, then the gradients remain unchanged, unlike the seasonal variations in the SST pattern. A comparison between projected SST-change in the HadCM3 model does not indicate that SST gradients in the Tropics will change significantly (not shown).
During the seasonal cycle, the region of warm water migrates north and south with the seasonal cycle of insolation, and the TC formation and tropical convection regions migrate with it. The areas of warmest SSTs, relative to surrounding regions, tend to have the more active convection than the cooler regions. One argument is that there are "winners and losers" in the tropics in terms of convection and that the warmer regions win out over cooler regions.  Physically, the overall control is provided not by absolute SST values, but by the large scale circulation (i.e., the rising and sinking regions of the Hadley and Walker Circulations) which through subsidence act to limit convection in the relatively cooler regions. However, it is also well-known that the atmosphere and the ocean are closely coupled in the tropics, giving rise to phenomena like the El Ni\~{n}o Southern Oscillation.

\section{Conclusion}

In summary, this paper does not attempt to provide a rigorous physical basis for the highly nonlinear SST-dependent behaviour. Instead, empirical evidence is analysed to assess the relationship between the number of TCs and the area of SST $>26.5$\degC.  It is not possible to even evaluate whether it would be likely to apply in the situation of large-scale SST warming. Nevertheless, a tentative projection has been made for the future, keeping in mind the caveats and uncertainties associated with such statistical extrapolations. The prediction of a rapidly increasing number of TCs after a certain area rests on the assumption that the statistical relations developed to model the seasonal variation of TC frequency can be extrapolated and used with the projected changes in future SSTs, but it is uncertain whether the statistical relation will hold for new conditions significantly different from the past. 
The relationship found here may be more complicated than it first appears, as other conditions also undergoing similar annual cycles may introduce misleading biases in the end results.
On the other hand, factors other than SST that may affect TCs are most likely not independent of the SSTs (shear, humidity, CAPE, El Ni\~{n}o, etc.), so that the area of the warm sea may also be regarded as proxy for all these aspects. The correlation analyses between predictions and year-to-year variations in the seasonal mean TC-number suggest that the statistical models capture a good part of the variations over the Atlantic and Caribbean basins, but not over the Indian ocean and the Pacific.

In concusion, these results are inconsistent with TCs being purely stochastic processes taking place over warm ocean regions. 
The statistical model may, however, explain the lack of a linear trend in $N$ in the past, despite a general increase in $A$. Furthermore, these results provide strong evidence for a real connection between $N$ and $A$ and against the claim that the region of cyclogenesis will not expand with the 26\degC\ isotherm. Thus, strong empirical evidence is presented here which suggest that the claim that the tropical cyclogenesis does not increases with the area enclosed by the 26\degC\ SST isotherm is false. These results furthermore suggest that there may be a non-linear relationship between the area of high SST and the number of TCs, which can explain why there has not yet been a clear upward trend in the number of TCs. 
\\

{\bf Acknowledgement}
Norsk Meteorolog Forening and the Norwegian Meteorological Institute. Much of this work was done during travels to Utrech (EMS), Berlin, and Helsinki.  
This manuscript has benefitted from reviewers from former submission to GRL/Climate Research - some of the arguments are acknoledged and met, whereas other comments were considered not convincing (despite coming from prominent experts within the tropical cyclone research community). Hence, this manuscript has benefitted from comments from several reviewers. Here, empirical data and relationships are presented, with the exception of the projection into the future (to be considered as an estimate of the upper bound for the number of TCs), and these empirical data/relationship ought to either support or falsify current hypotheses/theories, not vice versa. This principle is important, especially in the light of the current debate surrounding central issues regarding TCs, their relationship with global warming, or relationship with SSTs \citep{Scharroo2006,Sun2006,Michaels2005,Knutson2005,Pearce2005b,Smith2005}. It is acknowledged that the projection for the future is contriversial, however it is included to illustrate possible risks associated with uncertain knowledge about relationships between number of TCs and the ambient environment. 


\clearpage
\pagebreak

\bibliography{refs}

\clearpage
\pagebreak

\newpage
\end{document}